# Coulombically-stabilized oxygen hole polarons enable fully reversible oxygen redox

Iwnetim I. Abate[1,2], C. Das Pemmaraju[2], Se Young Kim[3], Kuan H. Hsu[1], Sami Sainio[4], Brian Moritz[2], John Vinson[5], Michael F. Toney[1,4], Wanli Yang[6], William E. Gent[1], Thomas P. Devereaux[1,2*], Linda F. Nazar[3*], William C. Chueh[1,2*]

[1]Department of Materials Science and Engineering, Stanford University, 496 Lomita Mall, Stanford, CA 94305, USA.

[2]Stanford Institute for Materials & Energy Sciences, SLAC National Accelerator Laboratory, 2575 Sand Hill Road, Menlo Park, CA 94025, USA.

[3]Department of Chemistry and the Waterloo Institute for Nanotechnology, University of Waterloo, 200 University Avenue West, Waterloo, Ontario N2L 3G1, Canada.

[4]Stanford Synchrotron Radiation Light source, SLAC National Accelerator Laboratory, 2575 Sand Hill Road, Menlo Park, CA 94025, USA.

[5]Material Measurement Laboratory, National Institute of Standards and Technology, Gaithersburg, Maryland 20899, USA.

[6]Advanced Light Source, Lawrence Berkeley National Laboratory, Berkeley, CA 94720, USA.

*email: tpd@stanford.edu (T.P.D.), lfnazar@uwaterloo.ca (L.F.N.), wchueh@stanford.edu (W.C.C.)

**Abstract**

Stabilizing high-valent redox couples and exotic electronic states necessitate an understanding of the stabilization mechanism. In oxides, whether they are being considered for energy storage or computing, highly oxidized oxide-anion species rehybridize to form short covalent bonds and are related to significant local structural distortions. In intercalation oxide electrodes for batteries, while such reorganization partially stabilizes oxygen redox, it also gives rise to substantial hysteresis. In this work, we investigate oxygen redox in layered $Na_{2-x}Mn_3O_7$, a positive electrode **material with ordered Mn vacancies. We show that coulombic interactions between oxidized oxide-anions and the interlayer Na vacancies can disfavor rehybridization and stabilize hole polarons on oxygen at 4.2 V vs. Na/Na$^+$. These coulombic interactions provide thermodynamic energy saving as large as O-O covalent bonding and enable ~ 40 mV voltage hysteresis over multiple electrochemical cycles with negligible voltage fade. Our results establish a complete picture of redox energetics by highlighting the role of coulombic interactions across several atomic distances and suggest avenues to stabilize highly oxidized oxygen for applications in energy storage and beyond.**

**Introduction**

Stable and reversible high-valent redox couples are foundational to (electro)chemical and catalytic transformations. A prominent application of such redox couples is intercalation battery electrodes.[1] In particular, the phenomenon of anionic redox in lithium- and sodium-ion positive electrodes has the potential to significantly improve cell energy density by providing additional high voltage capacity beyond that of most transition metal (TM) redox couples.[2–4] Two major hypotheses have guided the search for intercalation layered oxides with stable and reversible oxygen redox. First, short covalent bonding between oxygen atoms (e.g., ~ 1.5 Å O–O peroxo, vs. 2.8 Å in a typical TMO$_6$ octahedron)[4,5] and between TM and oxygen atoms (e.g., ~ 1.8 Å metal-oxo vs. 2.0 Å typical)[6] can stabilize oxidized oxide species against oxygen release. Second, the energetic penalty associated with the local distortion induced by such major bonding rearrangements can be mitigated by in-plane[7] and out-of-plane metal vacancy/antisite disorder[5,6,8,9] (i.e., so-called cation migration), leading to overall energy savings. The convergence of these two ideas has led to the current paradigm that oxygen redox and cation disordering come hand-in-hand to achieve stable but hysteretic redox in battery electrodes, typically involving hysteresis of several hundred mV.[9–11] As such, efforts have been directed at minimizing the hysteresis associated with transition metal hopping,[11] as well as identifying new structures that can accommodate local distortions induced by anion redox without disorder.[1] An alternative approach is to avoid such restructuring all together and therefore mitigate hysteresis.

Recently, several groups reported an anionic redox active material, $Na_{2-x}Mn_3O_7$, with a very small voltage hysteresis (~ 60 mV).[12–15] The negligible hysteresis is maintained even upon deep deintercalation unlike other materials where the magnitude of voltage hysteresis scales with the extent of deintercalation.[16] However, the mechanism that allowed such remarkable performance is still unclear. For example, Yamada and colleagues observed the emergence of a new absorption feature at ~531 eV upon anionic redox in X-ray absorption spectroscopy (XAS) measurements,[12] a feature resulting from a redox mechanism typical in oxygen-redox electrodes with large voltage hysteresis, such as $Li_{1+x}Ni_{1-y-z}Mn_yCo_zO_3$ (NMC),[9] $Li_{2-x}Ir_{1-y}Sn_yO_3$,[6] and $Na_{0.75}Li_{0.25}Mn_{0.75}O_2$.[8] However, since $Na_{2-x}Mn_3O_7$ and the other electrode materials exhibit differing hysteresis, distinctive redox mechanism and therefore XAS features would be expected. Therefore, an understanding of the redox mechanism in NMO could inform ways we can achieve similar exceptional behavior in other Na and Li based oxide positive electrode materials.

In this work, we report unambiguous experimental and computational spectroscopic confirmation of redox mechanism in NMO that is different from anionic redox positive electrode materials with high-hysteresis. We discover oxygen hole polaron formation and its electrochemical redox stability upon charging. The persistent ordered Mn vacancies in this electrode material provides the basis for calculating and understanding redox energetics in the absence of metal cation disordering. A clear picture for oxygen redox emerges: electrostatic interactions between oxygen hole polarons and Na vacancies compete with covalent bonding and transition metal disorder to determine the redox pathway. Through this understanding, we demonstrate that coulombic interactions can stabilize the otherwise unstable oxygen holes and provide energy saving as much as O-O covalent bonding, achieving exceptionally low, ~ 40 mV voltage hysteresis over electrochemical cycling with negligible voltage fade. Beyond batteries, stabilizing unusual polaronic states has many implications for defect engineering in condensed matter systems.

**Results and discussion**

Structurally, one out of every seven Mn sites are vacant in the TM layer of $Na_{2-x}Mn_3O_7$ ($Na_{4/7-x}Mn_{6/7}\square_{1/7}O_2$ where $\square$ = vacant sites in the TM layer) (Fig. 1A). This results in two oxygen sublattices: oxygen anions with two Mn neighbors (O-Mn2, termed "O1") and oxygen anions with three Mn neighbors (O-Mn3, termed "O2"). The substantial difference in ionic radii between octahedral $Mn^{IV}$ (0.53 Å) and $Na^I$ (1.02 Å) and the large interlayer spacing further increases the penalty for antisite defects. Previous works showed low-voltage hysteresis by charging up to 4.7 V.[12–14] However, despite the stable structure, voltage and capacity fading is observed > 4.4 V possibly due to side reactions and electrolyte decomposition (Fig. 1B and C). Since we are interested in determining the exact anionic redox mechanism, we cycled the material in the voltage region where electrolyte decomposition and surface Mn oxidation are minimal (fig. S1). Figure 1D shows the reversible electrochemistry of $Na_{2-x}Mn_3O_7$ between X = 0 and ~ 0.5 (25% desodiation) over 30 cycles at a C/20 rate between 3.5 V and 4.3 V vs $Na/Na^+$, where C is 166 mA g$^{-1}$. The symmetric and overlapping profiles of the differential capacity curve (Fig. 1D inset) further confirms reversibility of the redox process and absence of voltage fade in the voltage plateau region. This plateau has been attributed to de/intercalation of the octahedrally coordinated Na2 ions.[13] The Na1 ions that occupy trigonal prismatic sites in the layers above and below the Mn vacancies (Fig. 1A) are retained in this process. Importantly, the electrode exhibits a low voltage hysteresis of ~40 mV between charge and discharge, smaller by several factors than that reported in other known Li- and Na-based anionic-redox-active electrodes.[8–10] The voltage hysteresis remains largely constant with cycling (fig. S2).

The crystal structure is stable up to 4.3 V with < 0.5 Na remaining in the interlayer gap, unlike other Na layered TM oxides employing high-valent redox, such as α-$NaFeO_2$.[17] Structural stability over repeated redox cycles was confirmed by synchrotron powder X-ray diffraction (XRD, Fig. 1E) and Mn *K*-edge extended X-ray absorption fine structure (EXAFS, Fig. 1F). The Fourier-transformed EXAFS spectra, shown in Fig. 1F, reveal that a slight increase in the Mn-O bond length (< 2%) is the only local structural change upon charging to 4.3 V (Fig. 1F). This indicates that the bond order does not change significantly upon oxidation. Likewise, there is negligible change in the pre-edge feature in the Mn *K*-edge X-ray absorption spectrum, proving the absence of Mn migration to tetrahedral sites and the stability of the local structure (fig. S3D). The invariant XRD patterns shown through the 10$^{th}$ cycle in Fig. 1E indicate structural stability that is consistent with previous reports.[12–14]

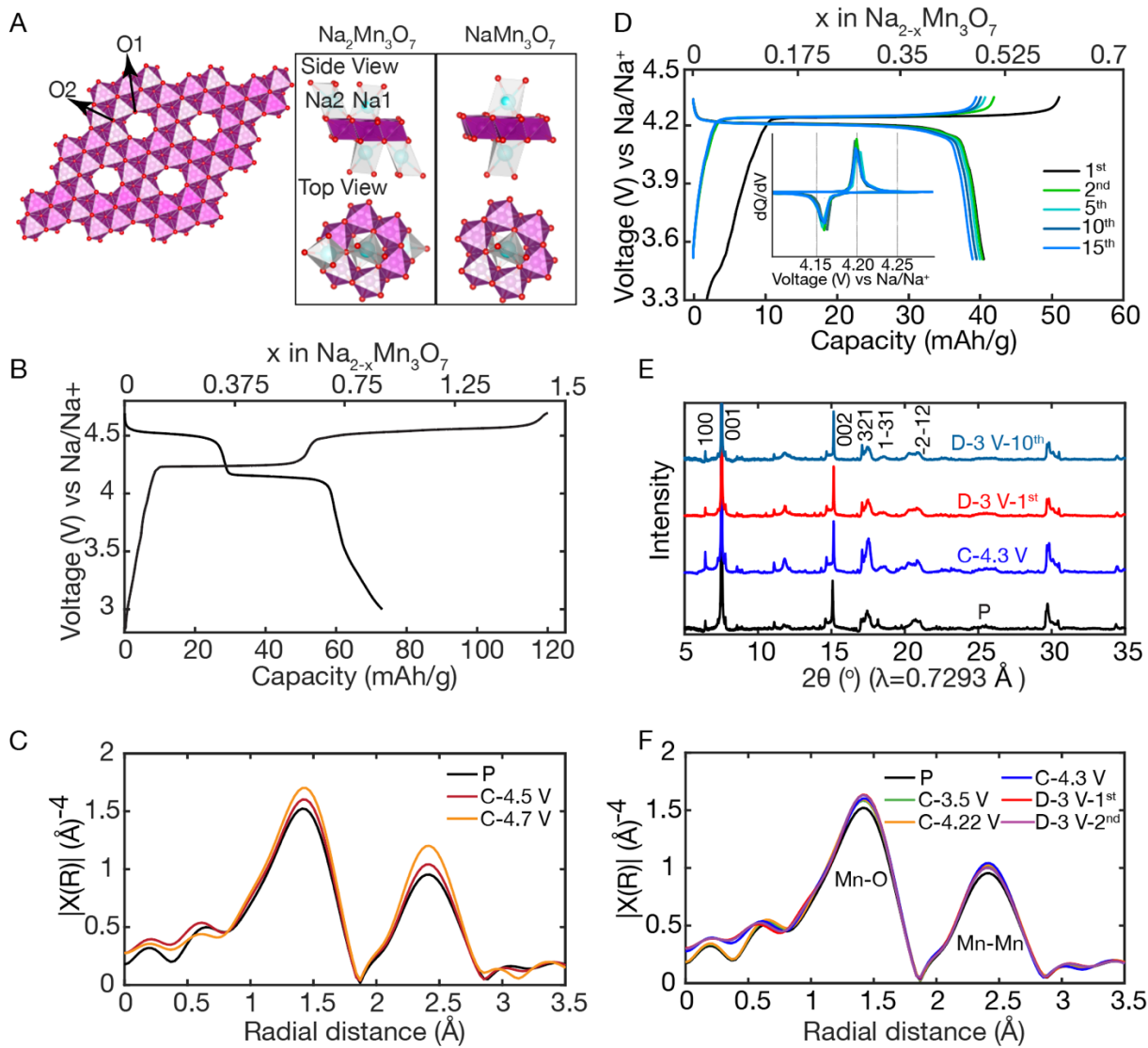

**Fig. 1. Reversible electrochemistry and stable structure.** (**A**) Structure of the Mn-O layer in Na$_2$Mn$_3$O$_7$ (Mn atoms shown in magenta and O atoms in red). One out of seven Mn sites are vacant, which creates two unique O environments: O–Mn2 (O1) in the ring surrounding the Mn vacancy; and O–Mn3 (O2). Top and side view of pristine and charged state. There are two sodium sites: Na2 (distorted octahedral) and Na1 (prismatic). (**B**) Voltage versus specific capacity at C/20 between 2.7 V and 4.7 V. (**C**) Fourier transform of the ex-situ EXAFS spectra taken at the same voltage conditions. The negligible change in the EXAFS profile indicates preservation of the local structure upon deep desodiation. (**D**) Voltage profile up to the 15$^{th}$ cycle at C/20 (1 C is equivalent to 166.2 mAh/g) between 3.5 V - 4.3 V vs Na/Na$^+$, with very low voltage hysteresis (~40 mV). Inset: differential capacity as a function of voltage (dQ/dV vs V). (**E**) Fourier transform of the ex-situ EXAFS spectra of the pristine material (P) and electrodes charged to 3.5 V, 4.22 V, and 4.3 V in the 1$^{st}$ cycle and discharged to 3 V in the 1$^{st}$ and 2$^{nd}$ cycle (D-3V-1$^{st}$ and D-3V-2$^{nd}$), respectively. The first peak corresponds to Mn-O bonds and the second to Mn-Mn bonds. The negligible change in the EXAFS profile indicates preservation of the local structure upon cycling. (**F**) Synchrotron powder X-ray diffraction of the pristine material; electrodes were charged to 4.3 V in the 1$^{st}$ cycle and discharged to 3 V at the 1$^{st}$ and 10$^{th}$ cycle (D-3V-1$^{st}$ and D-3V-10$^{th}$). The XRD patterns exhibit negligible changes that indicate minimal structural modification upon cycling (low angle XRD is shown in fig. S3).

To elucidate the nature of this highly reversible redox couple in Na$_{2-x}$Mn$_3$O$_7$, we performed X-ray absorption spectroscopy (XAS) at the O *K*-edge (Fig. 2) and at the Mn *L*- and *K*-edges (fig. S4-5). From total electron yield XAS, we confirmed a slight oxidation on the surface where we estimated the Mn oxidation state to change from ~ +3.9 in the pristine state to ~ +4.0 upon charging, fig. S4. This minimal oxidation contributes to the capacity only up to ~ 4.2 V (in the sloping region of the voltage curve). Both Mn *K* and *L*-edge spectra do not change during the voltage plateau. Rather, the plateau capacity at ~ 4.2 V arises from the oxidation of oxygen anions. The O *K*-edge resonant inelastic x-ray scattering (RIXS) map, Fig. 2A, shows the emergence of a new feature at an excitation energy of ~ 527.5 eV upon charging, which disappears on discharge. This new feature also appears in the XAS at the same energy as shown in Fig. 2B. The lack of change in the Mn *K*-edge (fig. S5) indicates that bulk Mn redox does not contribute to the capacity. Crucially, the oxygen redox feature remains largely constant in magnitude and reproducible between 1$^{st}$ and 10$^{th}$ cycle (Fig. 2C), directly confirming the reversibility of this oxygen redox couple. Upon deep desodiation up to 4.7V (50% desodiation) (Fig. 2D), we observed increase in the intensity of the ~527.5 eV peak, unlike previous reports (*14*).

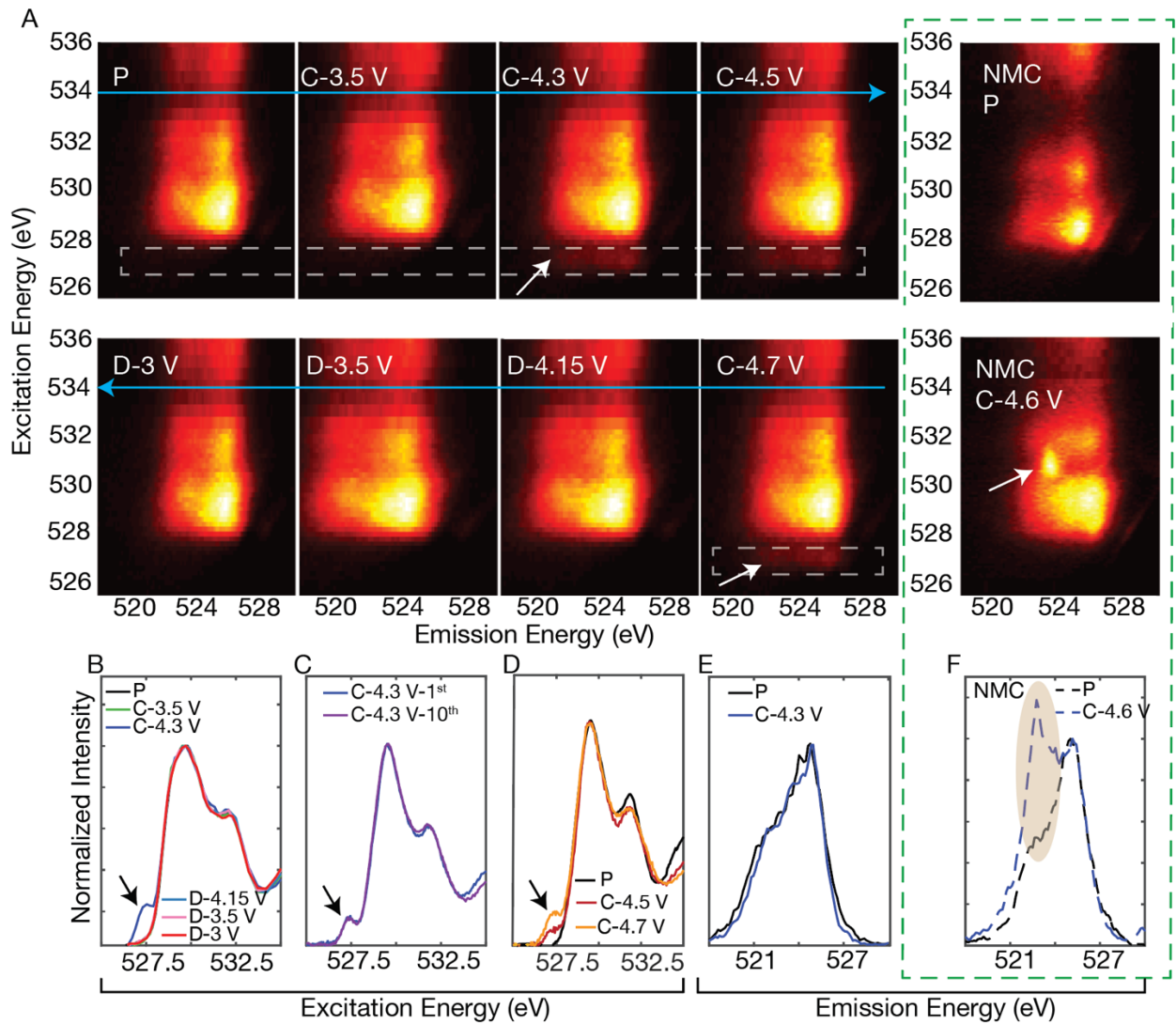

**Fig. 2. Oxygen-redox in Na$_{2-x}$Mn$_3$O$_7$.** (A) O $K$-edge RIXS maps of Na$_{2-x}$Mn$_3$O$_7$ from pristine to 4.7 V during charging (P to C-4.7 V) and from 4.15 V to 3 V during discharging in the bottom panel (D-4.15 V to D-3 V). The anionic-redox feature emerges at an excitation energy of ~527.5 eV upon charging (4.3-4.7V) as indicated by the white arrow. O $K$-edge RIXS maps of pristine and charged LR-NMC are also shown for comparison (NMC-P and NMC-4.6 V). The anionic-redox feature emerges at an excitation energy of ~531 eV upon charging (4.6 V) as indicated by the white arrow. All maps are taken during the first cycle. (B) O $K$-edge XAS spectra of Na$_{2-x}$Mn$_3$O$_7$, taken at the same voltage conditions as in panel A, showing the emergence of the peak at ~527.5 eV. (C) Comparison of the O $K$-edge XAS after charging (4.3 V) for the 1$^{st}$ and the 10$^{th}$ cycle. The normalized intensity of the ~527.5 eV feature is equivalent for the two spectra, indicating the stability of the anionic-redox mechanism. (D) O $K$-edge XAS of Na$_{2-}$

$_x$Mn$_3$O$_7$ from pristine to 4.7 V during charging (P to C-4.7 V). The increase in the intensity of the 527.5 eV peak when charging between 4.5 V and 4.7 V indicates that anionic redox with similar mechanism takes place on the ~ 4.3 V and ~ 4.5 V plateau. Emission spectra obtained by integrating the RIXS maps along a cut at ~531 eV excitation energy to investigate the presence of O-O dimers in Na$_{2-x}$Mn$_3$O$_7$ (E) & NMC (F). A peak at 523 eV (shaded region) from such a line cut has been assigned previously to O-O dimer formation. [6,9] The absence of this peak in Na$_{2-x}$Mn$_3$O$_7$ indicates a lack of dimer formation.

The spectroscopic signature of oxygen redox in Na$_{2-x}$Mn$_3$O$_7$ is distinct from a vast majority of known anion-redox-active electrodes. For example, NMC,[9] Li$_{2-x}$Ir$_{1-y}$Sn$_y$O$_3$,[6] and Na$_{0.75}$Li$_{0.25}$Mn$_{0.75}$O$_2$ [8] share the same spectroscopic signature in XAS or RIXS (~531 eV feature in XAS or a sharp feature at 531 eV excitation energy and 523 eV emission energy in RIXS, Fig. 2A for NMC). As can be seen in Figure 2D and E, such a feature is absent in Na$_{2-X}$Mn$_3$O$_7$. In electrodes with a 523 eV emission energy feature, the substantial structural disorder and large voltage hysteresis upon anionic redox has been linked to the formation of ~ 1.5 Å O-O dimers, which was proposed through comparison with a library of peroxide reference compounds using RIXS[1] and Raman spectroscopy though the precise speciation is still under debate.[18–20] The absence of a peak at ~ 800 cm$^{-1}$ in the Raman spectrum of Na$_{2-x}$Mn$_3$O$_7$ (fig. S6) indeed shows that 1.5 Å O-O peroxo moieties do not form.[18] We note that Na$_{0.6}$Li$_{0.2}$Mn$_{0.8}$O$_2$ [8] exhibits both the RIXS feature corresponding to O-O as well as the feature at an excitation energy of 527.5 eV observed in this work. However, as would be expected in an electrode material exhibiting oxygen dimerization, it exhibits a large voltage hysteresis and significant voltage fade.

Next, we performed *ab-inito* calculations using density-functional theory (DFT) with a Heyd–Scuseria–Ernzerhof functional, the most accurate functional for the ground state, and the Bethe-Salpeter Equation (BSE), a state-of-the art method for the excited state to understand the spectroscopic signature and energetic stability of oxidized oxide species. While it is challenging to employ first-principles calculations to identify the lowest energy structure of disordered materials (especially in the oxidized state), Na$_{2-x}$Mn$_3$O$_7$'s structural robustness against disordering makes it ideal for accurate first-principles analysis. Figure 3A shows a schematic top view of the pristine structure (sodium atoms colored cyan), with the corresponding density-of-states (DOS) shown in Fig. 3B. The two oxygen sublattices, "O1" and "O2" (Fig 1A) exhibit a very different contribution to the DOS. Close to the Fermi level, the relative atomic contribution to the density-of-states follows the order O1 > Mn > O2. Desodiation (charging) involves the preferential participation of the O1 species.

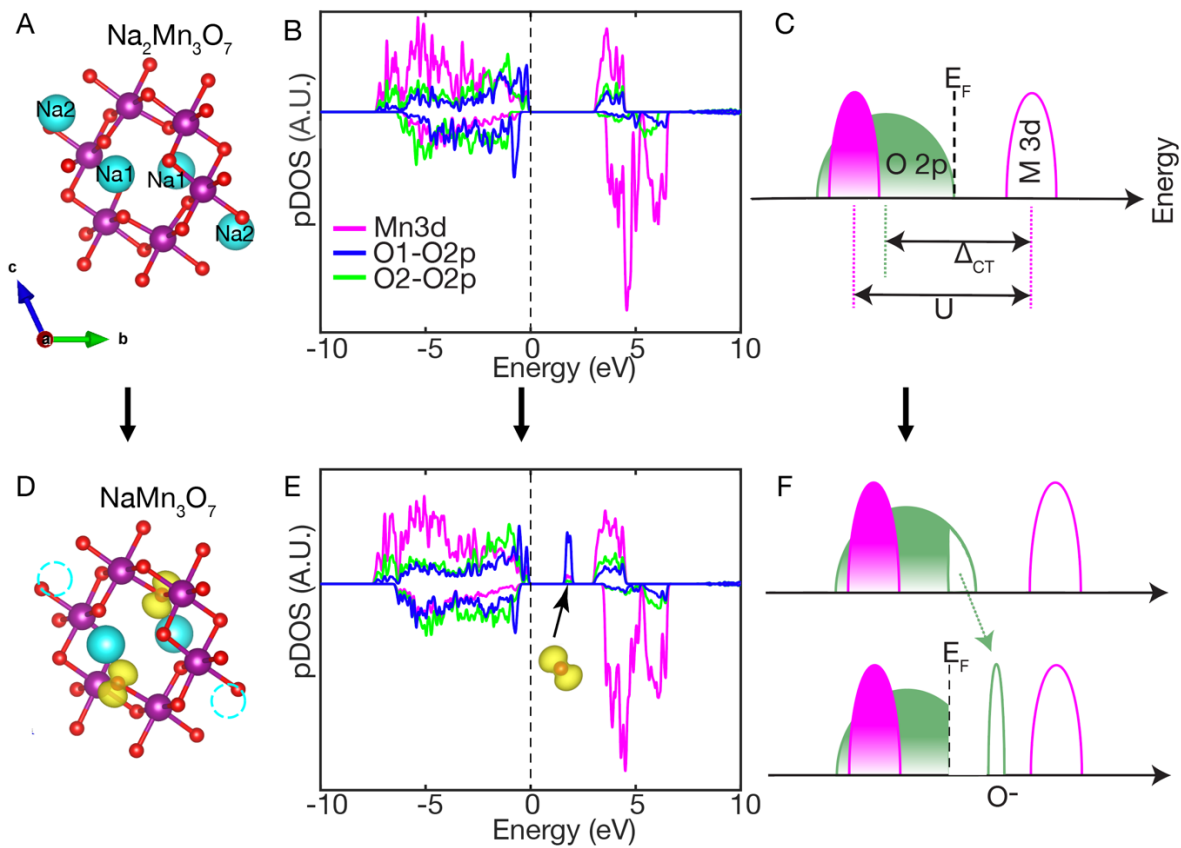

**Fig. 3. Density of states (DOS) of Na$_{2-x}$Mn$_3$O$_7$ from DFT.** (**A**) Structure in the pristine state (Na atoms in cyan). (**B**) Local DOS showing that O-$2p$ states dominate the band near the Fermi level, with O1 oxygen (blue) having a larger contribution to the DOS near the Fermi level than O2 oxygen (green). (**C**) Based on the Zaanen–Sawatzky–Allen (ZSA)[22] classification, Na$_{2-x}$Mn$_3$O$_7$ is a charge transfer insulator with the oxygen $2p$ band closer to the Fermi level than the metal (Mn) $3d$ band (Fig. 3C). Here, U > $\Delta_{CT}$, where U is the coulomb interaction energy for the $3d$ Mn electrons and $\Delta_{CT}$ is the charge transfer energy. (**D**) Structure after charging (desodiation, Na$_1$Mn$_3$O$_7$), with hole polarons on oxygen (spin density shown in yellow dumbbells), which form on the O1 species near the removed sodium atomic positions. In this work, the spin density of the hole polarons depicted is calculated over an energy range spanning the single particle level in the band gap associated with the species. (**E**) Local DOS in the charged

(desodiated) state, where hole polarons localize on O1 atoms to form in-gap states. (**F**) DOS schematic of charged state ($NaMn_3O_7$). Localized hole states form mid-gap states.

We simulated desodiated $NaMn_3O_7$ in the ground state by removing Na from the octahedral sites, and examined multiple configurations with holes localized on O1 and O2. Localized holes on O2 sites could not be realized following DFT self-consistent field iterations, suggesting that such a configuration is not a metastable state of the desodiated system. Instead, we find holes are localized on the under-coordinated O1 sites that surround the structural Mn vacancy (Fig. 3D). The local environment of O1 satisfies the condition for oxygen redox proposed by Seo et al.[21] First, we present results on the oxidation of two O1 located on opposite ends of the oxygen network surrounding a Mn vacancy (we term this the third-nearest neighbor, 3NN, configuration). The Mn-O1 bond length increases slightly, by < 2% (consistent with EXAFS, fig. S3). Projected density of states indicates that oxidation localizes and upshifts the O *2p* state at the top of the valence state by 1.4 eV (Figs. 3E & F), confirming that state corresponds to a oxygen hole polaron ($O^-$). We attribute the peak appearing at ~527.5 eV in the O *K*-edge XAS and RIXS to this localized hole polaron. Further confirmation is provided in the next section. We note that previous studies on this electrode material attributed the stability of O holes to hybridization between localized O *2p* and Mn $t_{2g}$ orbitals, but did not compare the relative energetics of O hole configurations and covalent bonded species.[12]

Next, we theoretically model the O *K*-edge XAS spectra using BSE method.[23,24] As shown in Figs. 4A and 4B, the experimentally-observed spectra in Fig. 2 are well reproduced. We can now assign the spectral features to O1 and O2. In the pristine material (Fig. 4A), the main XAS peak at 529.5 eV is attributed to equal contributions from O1 (blue) and O2 (green), corresponding to the $O_{1s}\rightarrow(Mn_{3d}-O_{2p}*)$ transitions. However, in the desodiated state (Fig. 4B), the peak at 527.5 eV arises from localized hole polarons ($O^-$) that form almost exclusively on O1 oxygen sites as shown in more detail in the inset of fig. S7 and Fig. 4B at 25% and 50% of desodiation, respectively. In general, these observations are in excellent agreement with the experimental XAS and RIXS results, as well as the projected DOS obtained from DFT.

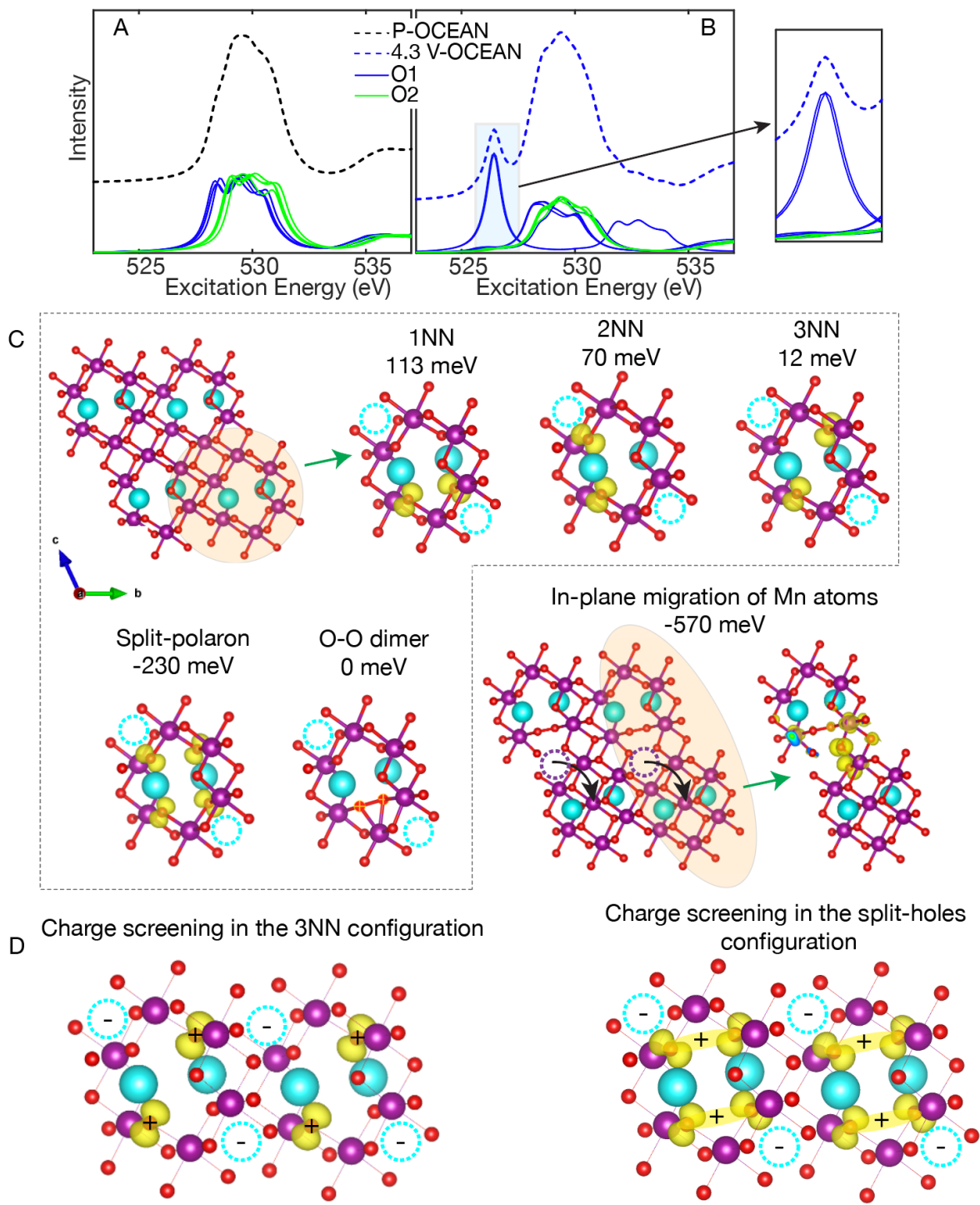

**Fig. 4. Confirmation via XAS simulation and O⁻ species stabilization mechanism.** (**A**) & (**B**) Simulated O *K*-edge XAS for pristine and charged samples (50% desodiation, $Na_1Mn_3O_7$), respectively. The dominant peak ~529.5 eV is from both O1 and O2 species; however, the O1 atoms have an out-weighted contribution to the localized hole polaron,

which gives rise to the peak at ~527.5 eV. (**C**) Comparing the energy (per Na atom removed) of different oxygen hole configurations. The split-polaron configuration (where a hole is shared among two O1 atoms) is the most stable configuration compared to first, third nearest neighbor (1NN & 3NN) and peroxo configurations when there is no in-plane Mn atom migration. However, the structure with in-plane migration (dotted circles in magenta and black arrow) is the most stable configuration. (**D**) The mechanism for the hole polarons. The split-polaron configuration provides a stronger screening for the long-range electrostatics between the Na vacancy sites (dotted circles in cyan) compared to the other configuration.

We note that holes on oxygen have been proposed in many systems.[8,12,21,25,26] Intensity changes in the ~531 eV region of the O *K* edge XAS (i.e. considerably above the absorption onset) are often claimed as evidence of such species, due to presumed large shifts in core level energies arising from the reduced electron density around oxidized O1 anions compared to $O^{2-}$. However, we have ruled out the presence of major chemical shifts (fig. S8 and the Supplementary Information for details), and show that localized hole polarons in fact manifest spectroscopically as an intensity at ~ 527.5 eV. We performed x-ray photoemission spectroscopy (XPS) simulation of charged state of $Na_{2-x}Mn_3O_7$. The relative chemical shift of the O-*1s* core-level on $O^-$ sites is estimated to be approximately 2 eV. Given that the $O^-$ hole mid-gap state has a single-particle energy that is roughly 1 eV below the conduction band (CB) minimum (see Figure 3F), and taking into account an additional core-excitonic interaction of roughly 1-1.5 eV in XAS, the $O^-$ *1s* → *2p* (hole) XAS transition is expected to appear as a pre-edge near ~527.5 eV to the $O^{2-}$ *1s* →CB XAS transitions that form the main edge at 529.5 eV. Our results highlight the importance of corroborating ground state DFT calculations with explicit spectroscopic simulation. It becomes increasingly clear that unoccupied O *2p* character states can only appear far above the absorption onset at the O *K* edge (i.e. ~ 531 eV) due to major covalent bonding rearrangements, such as the formation of < 1.5 Å O–O dimers, which raise the energy of the non-bonding O *2p* states by forming hybridized antibonding states.[1,6,8,9]

Driven by the significant covalent energy, oxygen hole polarons generally form O-O dimers.[4,5] Previous direct detection of O⁻ species in metal oxides by electron paramagnetic resonance has been reported only at liquid-helium temperatures after irradiation or chemical doping.[27] Some even argued O⁻ ions with negligible covalence cannot exist as bulk oxidized species,[28] and indeed there have been incorrect assignments of spectroscopic signals.[8,12,25,26] However, in $Na_{2-x}Mn_3O_7$ we observe the hole polaron XAS signal in the desodiated state for at least 10 cycles (Fig. 2C). To understand the stabilization mechanism, we systematically explored different arrangements involving two oxygen holes and two Na vacancies. In addition to the 3NN arrangement already considered, we now examine the oxidation of first and second-nearest neighbor O1 (1NN and 2NN, respectively (fig. S9)). Energetic stability is referenced to the peroxo dimer configuration (Fig. 4C), though more oxidized oxygen could be more stable (which we discussed later). For the 1NN configuration, it is 0.1 eV (per Na removed) less stable than the peroxo dimer. On the other hand, the 3NN configuration, wherein the two holes are maximally separated within the vacancy center (forming a linear chain consisting of oxygen polaron, Mn vacancy and oxygen polaron) is essentially as stable as the peroxo dimer. Given the highly localized nature of the oxygen holes, the stability of the 3NN relative to 1NN and peroxo dimer is dominated by coulombic interactions. Specifically, the 3NN configuration has a higher degree of symmetry, with the positively-charged oxygen holes providing better charge screening for the negatively-charged Na vacancies (See Fig. 4D).

In our calculations, we also identified an unusual split-hole polaron configuration, where a hole is shared among two O1 atoms via a very weak covalent bond (evidenced by the long O-O distance ~ 2.6 Å, as well as by the DOS (Fig 4D, fig. S10). This species can be loosely regarded as $O_2^{3-}$. This is unusual because cation vacancy bound deep acceptor states in wide gap oxides typically tend to form small polarons with spin density localized on a single O⁻ site.[29] We found that the split-hole configuration is the most stable: 0.23 eV (per Na removed) more stable than peroxo dimers and 3NN when no in-plane migration is considered. The explanation for this surprising stability (Table S1) lies in the selective removal of Na ions at the Na2 octahedral sites. Recall that Na ions at Na1 are not extracted during the 4.25 V plateau.[13] The positively charged holes are bound by coulombic forces to the negatively charged Na2 vacancy site. In the case of the split-hole polaron, a hole density is shared among two oxygen atoms and the oxidized O1 comes closer to the Na vacancy than in the 3NN configuration (Fig. 4D). The energy lowering via coulombic attraction in this state comes at the cost of some electron-lattice strain energy and exchange energy which tends to favor localization on an O1 site as in the 3NN case. We note that the XAS simulation for 3NN and split oxygen-hole polarons give similar spectroscopic signatures (peak ~ 527.5 eV) (fig. S10). Therefore, we cannot distinguish the two types of polarons within our spectroscopy simulation accuracy limits. Nonetheless, both configurations are predicted to be as or more stable than the peroxo dimer, which likely explains the complete absence of the latter and the minimal hysteresis.

To confirm the crucial role of coulombic interaction in stabilizing the oxygen hole polaron, we investigated several other extent and configuration of desodiation. Even at 25% desodiation, for an ordered Na2 vacancy arrangement, the split-hole polaron is predicted to be more stable than the localized hole by 0.3 eV (fig. S11). The situation is different, however, in the case of desodiating prismatic Na1 sites that sit on top of the Mn vacancy (fig. S12). In this case no split-hole polaron is predicted and a 3NN configuration wherein each hole localizes on an O1 site closest to one of the Na1 vacancies is stabilized. While Na1 desodiation is overall energetically unfavorable, the hole density distribution is consistent with an electrostatically driven mechanism (fig. S12). At different states of desodiation, the Na2 vacancies can be disordered in the material and different Mn vacancy sites can locally experience variations in sodium vacancy ordering. Nevertheless, since electrostatic forces in this material favor keeping hole densities separated, the dimer configuration will continue to be relatively disfavored (Fig 4, fig S10). Even in the presence of Na2 vacancy inhomogeneity, we expect Mn vacancy sites overall to support a distribution of the split-hole and localized O$^-$ hole configurations.

Recently, Bruce and colleagues suggested a new redox mechanism to describe anionic redox materials with high voltage hysteresis.[7,8] The mechanism involves in-plane TM migration and formation of O-O dimer with a bond length of ~ 1.2 Å. We explored this possibility in 50% desodiation of $Na_{2-x}Mn_3O_7$ (Fig 4C) and found that such configuration is more stable than split-hole polaron. This is consistent with observation where the hole polaron XAS feature disappears after more than 2 days from electrode harvesting, fig. S13). A previous EPR study also showed that hyperfine lines, which are related to the oxidized species in charged $Na_{2-x}Mn_3O_7$, disappeared after 7 days, also suggesting that the paramagnetic hole polarons ultimately form a diamagnetic O-O dimer.[13] These observations indicate that hole polarons in this material could have a substantial, but finite lifetime, potentially explaining why previous XAS reports on $Na_{2-x}Mn_3O_7$ did not observe the peak at 527.5 eV. However, we have shown that holes can reversibly dis/appear during dis/charge (Figs. 2B and 2C) for several cycles, as fresh hole polarons, created every cycle, do not dimerize on the time scale of our cycling condition (C/20), which is less than 48 hrs per cycle. Therefore, redox-structure decoupling and the exceptional reversibility remain possible in $Na_{2-x}Mn_3O_7$, since O-O dimers do not form during cycling, unlike other anionic redox active materials, possibly because of the effective polaron trapping and high migration barrier. Future work will investigate the stability of the hole polaron in the desodiated state.

**Conclusion**

Design rules around high-valent redox in intercalation electrodes have evolved around tuning (1) covalent bonding and (2) local structural distortion. Our experimental and theoretical investigation of $Na_{2-x}Mn_3O_7$ show that coulombic interactions is a crucial, third contribution to the overall redox energetics. In this system, significant coulombic interactions between oxygen hole polarons and the interlayer Na vacancy which occurs over several atomic distances, is sufficiently large to disfavor O-O dimerization within individual Mn vacancy sites restricting the space of possibilities for dimer formation. These species likely dominate in the growing class of systems where cation migration is inhibited during high valent redox. Coulombic interactions also rationalize previous hypotheses and computational predictions of partially oxidized, long (> 2 Å) O-O dimers with negligible TM hybridization forming during oxygen redox.[4] We suggest that tuning the in-plane cation-vacancy ordering, residual alkali content, and stacking order are means to control such coulombic effects. These tune the redox chemistry and oxidized species in layered positive electrode materials to give rise to low voltage hysteresis where cation disorder/migration is suppressed. This offers the possibility to tune structures to accommodate a higher hole polaron participation. The general concept of employing coulombic interactions to tune high valent redox chemistry can be extended to other structural motifs such as twists[30] and gating.[31] This approach could lead to new compositions and structures with improved stability in a highly oxidized state, which has applications in energy storage and beyond.

## Author contributions

I.I.A, W.E.G. and W.C.C. conceived the project; W.C.C., L.F.N., M.F.T and W.Y. supervised the experiments; T.P.D supervised the theoretical calculations; I.I.A and S.Y.K synthesized the materials and performed electrochemical measurements. I.I.A performed XAS, RIXS, EXAFS, Raman and XRD measurements. I.I.A, C.D.P and K.H.H. performed the theoretical calculations. I.I.A, C.D.P and S.Y.K analyzed and interpreted the data. S.S assisted in collection of XAS data. J.V. and B.M. assisted in interpreting the simulation results. I.I.A wrote the manuscript with inputs from all authors; W.C.C. directed the overall research.

## Conflicts of interest

There are no conflicts of interest to declare.

## Acknowledgments

The authors would like to thank Kipil Lim, Peter Csernica, Chunjing Jia, and Che-Ning Yeh for valuable discussions, and Loza Tadesse for Raman measurements. This work was supported by the U.S. Department of Energy (DOE), Office of Basic Energy Sciences, Division of Materials Sciences and Engineering (contract DE-AC02- 76SF00515). I.I.A. was additionally supported by the Stanford DARE fellowship program. W.E.G. was supported by BASF. L.F.N. would like to acknowledge NSERC for funding through their Discovery Grant platform, and the Canada Research Chair program. Use of the ALS was supported by the Office of Science, Office of Basic Energy Sciences, of the US DOE under contract no. DE-AC02-05CH11231. Use of the SSRL, SLAC National Accelerator Laboratory, was supported by the Office of Science, Office of Basic Energy Sciences, of the US DOE under contract no. DE-AC02-76SF00515. Part of this work was performed at the Stanford Nano Shared Facilities, supported by the National Science Foundation under award ECCS-1542152. The computational work used resources at the

**Notes and references**